\documentclass[12pt]{iopart}
\usepackage{epsfig}
\usepackage{graphicx}

\newcommand{\room}{\rule[-0.3cm]{0cm}{0.8cm}}
\newcommand{\be}{\begin{equation}}
\newcommand{\ee}{\end{equation}}
\newcommand{\bd}{\begin{displaymath}}
\newcommand{\ed}{\end{displaymath}}
\newcommand{\vsp}{\vspace*{3mm}}

\newcommand{\bra}{\langle}
\newcommand{\ket}{\rangle}
\newcommand{\order}{{\cal O}}

\newcommand{\minus}{{\!-\!}}
\newcommand{\sgn}{{\rm sgn}}

\newcommand{\bsigma}{{\mbox{\boldmath $\sigma$}}}

\newcommand{\bxi}{{\mbox{\boldmath $\xi$}}}

\begin{document}

\title{Finite Connectivity Attractor Neural Networks}

\author{B Wemmenhove$^\dag$ and A C C Coolen$^\ddag$}
\address{$\dag$ Institute for
Theoretical Physics, University of Amsterdam, Valckenierstraat 65,
1018 XE Amsterdam, The Netherlands\\ $\ddag$ Department of
Mathematics, King's College, University of London, The Strand,
London WC2R 2LS, United Kingdom}

\begin{abstract}
We study a family of diluted attractor neural networks with a
finite average number of (symmetric) connections per neuron. As in
finite connectivity spin glasses,  their equilibrium properties
are described by order parameter functions, for which we derive an
integral equation in replica symmetric (RS) approximation. A
bifurcation analysis of this equation reveals the locations of the
paramagnetic to recall and paramagnetic to spin-glass transition
lines in the phase diagram. The line separating the retrieval
phase from the spin-glass phase is calculated at zero temperature.
All phase transitions are found to be continuous.
\end{abstract}

\pacs{75.10.Nr, 05.20.-y, 64.60.Cn}

\ead{\tt wemmenho@science.uva.nl, tcoolen@mth.kcl.ac.uk}

\section{Introduction}

Spin models of recurrent neural networks have been studied
intensively within equilibrium and non-equilibrium statistical
mechanics, especially after Hopfield \cite{Ho82} emphasized their
link with spin-glasses. In Hopfield's picture, the free energy
minima of glassy systems (fixed-point attractors of the dynamics)
are given an information processing interpretation: in recurrent
neural networks they represent stored items of information
(`patterns'), whose locations in phase space are the result of
suitable modification of the neuron interactions (`learning').
Spin glass theory, especially the replica method \cite{MPV87}, was
shown to be an efficient tool with which to study the equilibrium
properties of Hopfield type  models (or `attractor neural
networks')  with full connectivity \cite{AGS85a,AGS85b}. It was
clear from the start that full connectivity is an undesirable
simplification of biological reality, made only for mathematical
convenience. However, it was also clear that solving neural
network models with restricted range interactions in finite
dimensions $D$ was either pointless (no phase transitions for
$D=1$) or impossible (for $D>1$). This dilemma prompted the study
of so-called diluted models, where for each spin all but a
randomly selected subset of size $c$ of interactions are removed;
such can be done either while preserving interaction symmetry, or
asymmetrically (in the latter case detailed balance no longer
holds). Different regimes for the scaling of $c$ with the system
size $N$ were found to give different physics, and to require
different mathematical techniques. The early models of
\cite{AGS85a,AGS85b} correspond to $c=N$.  The Hopfield model with
so-called extreme dilution (i.e.
$\lim_{N\to\infty}c^{-1}=\lim_{N\to\infty}c/N=0$, such as $c=\log
N$) was studied in \cite{DGZ} for the case of fully asymmetric
dilution, and in  \cite{WS91} for the case of symmetric dilution.
Both in fully connected and in symmetrically extremely diluted
networks close to saturation one finds a conventional replica
theory \'{a} la \cite{MPV87}, with the familiar replica order
parameter matrix. An alternative route away from full
connectivity, which preserves both the potential for phase
transitions and solvability, was followed in
\cite{Skantzos1,Skantzos2} (the $1+\infty$ dimensional attractor
networks). For recent reviews of the equilibrium and
non-equilibrium statistical mechanics of recurrent neural networks
see e.g. \cite{CH14,CH15}.

In this paper we study the as yet unsolved class of symmetrically
diluted attractor neural networks with {\em finite connectivity},
where the average number $c$ of bonds per neuron is finite,
independent of the system size $N$. For bond-disordered spin
systems, the finite connectivity regime (which so far has been
addressed only in the context of spin glasses
\cite{VB85,KS87,MP87,DM87,WS88,Mo98}, error correcting codes
\cite{codesrefSaad,codesrefNishimori} and satisfiability  problems
\cite{sat1,sat2,sat3,sat4}) requires order parameter functions,
which generalize the replica matrices of \cite{MPV87}. For finite
$c$ the replica symmetry breaking theory (RSB) is still under
development \cite{WS88,DG89,LG90,GD90,Mo98,PT02}. Here we apply
the finite connectivity spin glass replica techniques to a general
family of attractor neural networks (which includes the Hopfield
model, but also the so-called clipped Hebbian synapses, as in e.g.
\cite{LvH}), within the replica symmetry ansatz (RS). We obtain
phase diagrams in the $(\alpha,T)$ plane, for arbitrary finite
$c$, where $\alpha=p/c$ ($p$ giving the number of stored patterns)
and $T$ is the temperature. These diagrams contain a paramagnetic
phase (P), a recall phase (R), and a spin-glass phase (SG), all
separated by second order transitions. We also analyze the RS
ground state, and show how for $c\to\infty$ and arbitrary $T$  one
recovers the simpler results of \cite{WS91}. The most surprising
outcome of our calculations is the low values of the connectivity
$c$, only barely exceeding the percolation threshold, which are
required for the system to operate effectively as an attractor
neural network; equivalently, the robustness of such information
processing systems against excessive dilution and/or physical
damage.

\section{Model Definitions}

We study Ising spin neural network models, with microscopic states
defined by the $N$-neuron state vector
$\bsigma=(\sigma_1,\ldots,\sigma_N)\in\{-1,1\}^N$. Here
$\sigma_i=1$ if neuron $i$ fires, and $\sigma_{i}=-1$ if it is at
rest. Upon imposing standard Glauber-type dynamics, where the
spins align stochastically to local fields of the form
$h_i(\bsigma)=\sum_{j\neq i}J_{ij}\sigma_j$ with symmetric
interactions  $\{J_{ij}\}$, these systems will evolve to
thermodynamic equilibrium,  described by the Hamiltonian
$H(\bsigma)=-\sum_{i<j}^N J_{ij}\sigma_i\sigma_j$ and the
associated free energy
\be
F=-\beta^{-1}\log Z,~~~~~~~~Z=\sum_{\bsigma}e^{-\beta H(\bsigma)}
\ee The interactions are defined as a diluted and generalized
version (in the spirit of \cite{LvH}) of the standard Hebbian
recipe, with $c_{ij}\in\{0,1\}$ and $c_{ij}=c_{ji}$ for all
$(i,j)$:
\be
J_{ij}=\frac{c_{ij}}{c}~\phi\left(\sum_{\mu=1}^p\xi_i^\mu\xi_j^\mu
\right) ~~~~~~~~~~ \xi_i^\mu\in\{-1,1\}~~{\rm  for~ all}~ (i,\mu)
\label{eq:Jij} \ee The $p$ vectors
$\bxi^\mu=(\xi_1^\mu,\ldots,\xi_N^\mu)\in\{-1,1\}^N$ represent
patterns to be stored in the system. The $c_{ij}\in\{0,1\}$ define
the connectivity of the network, and act as quenched disorder.
They are drawn randomly and independently from
\be
P(c_{ij})=\frac{c}{N}\delta_{c_{ij},1}+[1-\frac{c}{N}]\delta_{c_{ij},0}
\ee The average number of connections to any neuron is $c$.
Averages over the realizations of the $\{c_{ij}\}$ will be denoted
as $\overline{\cdots}$. The function $\phi(x)$ in (\ref{eq:Jij})
need not be specified at this stage; for $\phi(x) = x$ one returns
to the symmetrically diluted Hopfield  model, for
$\phi(x)=\sqrt{p}~\sgn(x)$ one finds diluted and clipped Hebbian
synapses, etc. The relevant and nontrivial
 scaling regime is that where $\lim_{N\to\infty}p/c=\alpha$,
with $0<\alpha<\infty$.

Models of the type (\ref{eq:Jij}) have so far been studied in
regimes where  $\lim_{N\to\infty}c^{-1}=0$. Here we study the more
complicated scaling regime where $c=\order(N^0)$. Our objective is
to solve the model by calculating the disorder-averaged free
energy per spin in the thermodynamic limit: $f=-\lim_{N\to
\infty}(\beta N)^{-1}\overline{\log Z}$.
 The
replica identity $\overline {\log Z}=\lim_{n\to 0}n^{-1}\log
\overline{Z^n}$ allows us to write this in the standard manner as
\begin{eqnarray}
f&=&-\lim_{N\to \infty}\lim_{n\to 0}\frac{1}{\beta
Nn}\log\sum_{\bsigma^1\ldots\bsigma^n}
\overline{e^{\frac{\beta}{c}\sum_{i<j}c_{ij}\phi(\bxi_i \cdot \bxi_j)
\sum_\alpha
\sigma^\alpha_i\sigma^\alpha_j}}
\label{eq:free_energy_def}
\end{eqnarray}
Due to $c=\order(N^0)$, one can expand this expression for
$N\to\infty$:
\be
\hspace*{-20mm} f =-\lim_{N\to \infty}\lim_{n\to 0}\frac{1}{\beta
Nn}\log\sum_{\bsigma^1\ldots\bsigma^n} \exp\left\{
\frac{c}{2N}\sum_{ij}\left[e^{\frac{\beta}{c} \phi(\bxi_i \cdot
\bxi_j)\sum_\alpha \sigma^\alpha_i\sigma^\alpha_j} \minus
1\right]+\order(N^0)\right\} \label{diso_aver_f_e} \ee In
non-diluted disordered spin systems, the sum in the exponent would
have been quadratic in the spin variables; there the free energy
can be linearized by a gaussian transformation,  leading to a
single spin problem with the conventional replica order parameter
matrix $\{q_{\alpha \beta}\}$ \cite{MPV87}. In systems with finite
connectivity, in contrast, one finds  more complicated order
parameters which encode  higher order correlations between
replicas \cite{VB85}. In these  models it is more convenient to
use an order parameter \emph{function} \cite{KS87,MP87,DM87},
describing the distribution of spin variables in the various
replicas, from which all the conventional order parameters can be
derived.

\section{Calculation of the Free Energy}

\subsection{Replica Analysis and Sub-lattice Order Parameters}

To work out (\ref{diso_aver_f_e}) we exploit the fact that for
$c=\order(N^0)$, also $p=\order(N^0)$. This allows us to use the
concept of sublattices \cite{HK86} (of which there are $2^p$):
\be
I_\bxi=\{i|~\bxi_i=\bxi\}~~~~~~~~~~ p_\bxi=|I_\bxi|/N
 \ee
We define $\bsigma_i=(\sigma_i^1,\ldots,\sigma_i^n)$, and
henceforth abbreviate averages over the sublattices as $\bra
f(\bxi) \ket_{\bxi} = \sum_{\bxi} p_{\bxi} f(\bxi)$.
 The trace
in (\ref{diso_aver_f_e}) can now be written as
\be
\hspace*{-15mm} \sum_{\bsigma^1\ldots\bsigma^n}
\exp\left\{\frac{c}{2N}\sum_{\bsigma \bsigma^\prime}\sum_{\bxi
\bxi^\prime}\left[e^{\frac{\beta}{c}\phi(\bxi\cdot\bxi^\prime)
\sum_\alpha \sigma_\alpha\sigma_\alpha^\prime} \minus
1\right]\sum_{i\in I_\bxi}\sum_{j\in I_{\bxi^\prime}}
\delta_{\bsigma,\bsigma_i}\delta_{\bsigma^\prime,\bsigma_j}+\order(N^0)\right\}
\label{eq:thetrace} \ee We next introduce a spin distribution
$P_{\bxi}(\bsigma)$ within each sublattice, with $\bsigma =
\{\sigma^1,\ldots, \sigma^n\}$, to be isolated upon inserting
suitable $\delta$-distributions into (\ref{diso_aver_f_e}):
\be
1=\int\!\prod_{\bxi
\bsigma}\left\{dP_\bxi(\bsigma)~\delta\left[P_{\bxi}(\bsigma)-
\frac{1}{|I_\bxi|} \sum_{i\in I_\bxi}
\delta_{\bsigma,\bsigma_i}\right]\right\} \label{eq:delta_rep}
 \ee
From these distributions, one derives the more familiar types of
observables such as replicated sublattice magnetizations
$m^\alpha_{\bxi}$ and replicated pattern overlaps $m^{\mu
\alpha}$:
\be
m^\alpha_{\bxi} = \sum_{\bsigma} P_{\bxi}(\bsigma) \sigma^\alpha
~~~~~~~~~~  m^{\mu \alpha} = \bra \xi^\mu
m^\alpha_{\bxi}\ket_{\bxi} \label{overlaps} \ee The introduction
of integral representations for the delta functions in
(\ref{eq:delta_rep}) generates conjugate variables
$\hat{P}_{\bxi}(\bsigma)$. The trace in (\ref{eq:thetrace})  is
now performed trivially, resulting in an integral to be evaluated
by steepest descent:
\begin{eqnarray}
\hspace*{-10mm} f &=&-\lim_{N\to \infty}\lim_{n\to
0}\frac{1}{\beta Nn}\log\int\!\left[\prod_{\bxi \bsigma}
dP_\bxi(\bsigma)d\hat{P}_\bxi(\bsigma)\right]\exp\left\{\room
iN\sum_{\bsigma}\bra\hat{P}_\bxi(\bsigma)P_{\bxi}(\bsigma)\ket_{\bxi}\right\}\nonumber
\\
\hspace*{-10mm} && \times \exp\left\{ \frac{cN}{2}\bra\bra
\sum_{\bsigma \bsigma^\prime}
P_\bxi(\bsigma)P_{\bxi^\prime}(\bsigma^\prime)
\left[e^{\frac{\beta}{c}\phi(\bxi\cdot\bxi^\prime)(\bsigma\cdot\bsigma^\prime)}
\minus 1\right]\ket\ket_{\bxi\bxi^\prime}+\order(N^0)\right\}
\nonumber
\\
\hspace*{-10mm} &&\times \exp\left\{ N\bra
\log\left[\sum_{\bsigma}
e^{-i\hat{P}_\bxi(\bsigma)}\right]\ket_{\bxi}\right\} \nonumber \\
\hspace*{-10mm} &=&
 -\lim_{n\to
0}\frac{1}{\beta n}{\rm extr}_{\{
P_\bxi(\bsigma),\hat{P}_\bxi(\bsigma)\}}~\left\{
i\sum_{\bsigma}\bra
\hat{P}_\bxi(\bsigma)P_{\bxi}(\bsigma)\ket_{\bxi}
 +\bra
\log\left[\sum_{\bsigma}e^{
-i\hat{P}_\bxi(\bsigma)}\right]\ket_{\bxi} \right.\nonumber
\\ \hspace*{-10mm} &&\left.~~~~~~~~~~
 + \frac{1}{2}c\bra\bra \sum_{\bsigma \bsigma^\prime}
P_\bxi(\bsigma)P_{\bxi^\prime}(\bsigma^\prime)
\left[e^{\frac{\beta}{c}\phi(\bxi\cdot\bxi^\prime)(\bsigma\cdot\bsigma^\prime)}
\minus 1\right]\ket\ket_{\bxi\bxi^\prime} \right\}
\label{eq:free_energy_further}
\end{eqnarray}
Variation with respect to  $\{P_{\bxi}(\bsigma)\}$ gives an
equation with which to  eliminate the conjugate order parameters,
resulting in
 $f=\lim_{n\to 0}{\rm
extr}_{\{P_{\bxi}(\bsigma)\}}f[\{P_{\bxi}(\bsigma)\}]$, where
\begin{eqnarray}
\hspace*{-10mm} f[\ldots] &=& \frac{c}{2\beta n} \bra\bra
\sum_{\bsigma \bsigma^\prime}
P_\bxi(\bsigma)P_{\bxi^\prime}(\bsigma^\prime)
\left[e^{\frac{\beta}{c}\phi(\bxi\cdot\bxi^\prime)(\bsigma\cdot\bsigma^\prime)}
\minus 1\right]\ket\ket_{\bxi\bxi^\prime} \nonumber \\
\hspace*{-10mm} && -\frac{1}{\beta n}\bra \log\left[\sum_{\bsigma}
\exp\left\{ c\bra \sum_{\bsigma^\prime}
P_{\bxi^\prime}(\bsigma^\prime)
\left[e^{\frac{\beta}{c}\phi(\bxi\cdot\bxi^\prime)(\bsigma\cdot\bsigma^\prime)}
\minus 1\right]\ket_{\bxi^\prime}\right\} \right]\ket_{\bxi}
\label{eq:free_energy_further2} \end{eqnarray} Further variation,
again with respect to $P_{\bxi}(\bsigma)$, then leads to the
saddle point equation
\begin{eqnarray}
P_{\bxi}(\bsigma)
=
\frac{ \exp\left\{c \bra \sum_{\bsigma^\prime}
P_{\bxi^\prime}(\bsigma^\prime)\left[ e^{\frac{\beta}{c} \phi(\bxi
\cdot \bxi^\prime)(\bsigma \cdot \bsigma^\prime)} -1
\right]\ket_{\bxi^\prime} \right\} } {
\sum_{\bsigma^\prime}\exp\left\{c \bra
\sum_{\bsigma^{\prime\prime}}
P_{\bxi^{\prime\prime}}(\bsigma^{\prime\prime})\left[
e^{\frac{\beta}{c} \phi(\bxi^{\prime\prime} \cdot
\bxi)(\bsigma^{\prime\prime} \cdot \bsigma^\prime)} -1
\right]\ket_{\bxi^{\prime\prime}} \right\} }
\label{eq:saddlepoint}
\end{eqnarray}
When deriving (\ref{eq:saddlepoint}) it first appears that one
must also allow for adding to $\{P_{\bxi}(\bsigma)\}$ any vector
in the kernel of the $2^{p+n}\times 2^{p+n}$ matrix
$U(\bxi,\bsigma;\bxi^\prime,\bsigma^\prime) =
\exp[\frac{\beta}{c}\phi(\bxi \cdot \bxi')(\bsigma \cdot
\bsigma')] -1$. However, adding such elements is seen to generate
new  expressions for $\{P_{\bxi}(\bsigma)\}$ which again solve
(\ref{eq:saddlepoint}); hence (\ref{eq:saddlepoint}) is always
satisfied at the saddle-point.

An alternative route for calculating the free energy will turn out
 to be more convenient for calculating zero temperature
properties, such as ground state energy and entropy. Here we first
vary (\ref{eq:free_energy_further}) with respect to
$\hat{P}_{\bxi}(\bsigma)$, resulting in
 $f=\lim_{n\to 0}{\rm
extr}_{\{P_{\bxi}(\bsigma)\}}f[\{P_{\bxi}(\bsigma)\}]$ subject to
the constraints $\sum_{\bsigma}P_{\bxi}(\bsigma)=1$ for all
$\bxi$, with
\begin{eqnarray}
f[\ldots]& =& -\frac{c}{2\beta n} \bra \bra
\sum_{\bsigma\bsigma^\prime} P_{\bxi}(\bsigma)
P_{\bxi^\prime}(\bsigma') \left[ e^{\frac{\beta}{c}\phi(\bxi \cdot
\bxi^\prime)(\bsigma \cdot \bsigma^\prime)}-1\right]
\ket\ket_{\bxi\bxi^\prime} \nonumber \\ &&
 +\frac{1}{\beta n} \bra
\sum_{\bsigma} P_{\bxi}(\bsigma) \log P_{\bxi}(\bsigma)
\ket_{\bxi} \label{eq:alt_free_e}
\end{eqnarray}
The conjugate order parameter functions $\hat{P}_{\bxi}(\bsigma)$
are now found to act as Lagrange multipliers, imposing the
normalization constraints on the distributions
$P_{\bxi}(\bsigma)$.

\subsection{RS Order Parameter Equations and Free Energy}

In order to take the $n\to 0$ limit  we make the replica symmetric
(RS) ansatz for the order parameters. One anticipates RS to be
broken in the low temperature phases \cite{Ho82,VB85,WS91}, but to
hold up to the first phase transition away from the paramagnetic
phase, as in the Viana-Bray model \cite{VB85}. In the present
context, RS is equivalent to assuming that the $P_{\bxi}(\bsigma)$
are invariant under permutations of the components of $\bsigma$,
i.e. only depend on the sum $\sum_{\alpha} \sigma^\alpha$. Upon
defining effective fields \cite{KS87}, viz. $h_i \equiv
\frac{1}{\beta}\tanh^{-1}\langle \sigma_i^\alpha \rangle_{\beta}
$,  our RS order parameter functions can thus be written
\cite{Mo98} as
\be
P_{\bxi}(\bsigma) = \int\! dh~ W_{\bxi}(h) \frac{e^{\beta h
\sum_\alpha \sigma^\alpha}}{[2\cosh(\beta h)]^n}
\label{eq:RSansatz} \ee
Here $W_{\bxi}(h)\geq 0$ and
$\int\!dh~W_{\bxi}(h)=1$. One can write sub-lattice magnetizations
and higher order observables in terms of the $W_{\bxi}(h)$:
\be
\sum_{\bsigma} P_{\bxi}(\bsigma)\sigma^\alpha \sigma^\beta \ldots
\sigma^r = \int\! dh~ W_{\bxi}(h) \tanh^r(\beta
h)~~~~~~(\alpha\!<\!\beta\!<\!\ldots \!<\!r)
\label{eq:orderparams} \ee The pattern overlaps $m^{\mu\alpha}$
are now independent of $\alpha$, $m^{\mu\alpha}=m_\mu$, where
\be
 m_{\mu} = \bra \xi^\mu
\int\! dh~ W_{\bxi}(h) \tanh(\beta h) \ket_{\bxi}
\label{eq:RSoverlaps} \ee
We substitute (\ref{eq:RSansatz}) into
(\ref{eq:saddlepoint}), and isolate the occurrences of
$\sum_\alpha\sigma^\alpha$ by inserting
\be
1 = \sum_{m=-\infty}^{\infty} \int_0^{2\pi} \frac{d\hat{m}}{2\pi}~
e^{i\hat{m}(m-\sum_\alpha \sigma^\alpha)} \ee After some mostly
straightforward manipulations\footnote{The only technical subtlety
is the need to take $n\to 0$ for {\em even} $n$, to avoid in the
denominator of (\ref{eq:saddlepoint}) tricky terms like
$\log\cos(\hat{m})$ with $\hat{m}<0$. The same issue arises when
calculating the RS free energy.} one can then take the $n\to 0$
limit and find (\ref{eq:saddlepoint}) converting into: \be
\hspace*{-21mm}
 \int\!dh~W_{\bxi}(h)e^{\beta h m} =
\exp\left\{ c~\bra \int\! dh^\prime
W_{\bxi^\prime}(h^\prime)\left[e^{m \tanh^{-1}[\tanh(\beta
h^\prime) \tanh( \frac{\beta}{c}\phi(\bxi \cdot
\bxi^\prime))]}-1\right]\ket_{\bxi^\prime}\right\}
\label{eq:saddle_a} \ee This
 is to hold for any real $m$.  Provided the various
integrals exist, we may now also put $m\to im/\beta$ and carry out
an inverse Fourier transform (following \cite{Mo98}), leading to
\begin{eqnarray}
W_{\bxi}(h) &=& \int\! \frac{dm}{2\pi} ~e^{-imh} \nonumber\\
&&\hspace*{-10mm}\times \exp\left\{ c~\bra \int\! dh^\prime
W_{\bxi^\prime}(h^\prime)\left[e^{\frac{im}{\beta}
\tanh^{-1}[\tanh(\beta h^\prime) \tanh( \frac{\beta}{c}\phi(\bxi
\cdot \bxi^\prime))]}-1\right]\ket_{\bxi^\prime}\right\}
\label{eq:W_saddlepoint}
\end{eqnarray}
These are the final equations from which to solve the RS order
parameters, i.e. the $2^p$ effective field distributions
$W_{\bxi}(h)$. \vsp

Upon inserting the RS ansatz (\ref{eq:RSansatz}) into  expression
(\ref{eq:free_energy_further2}) for the free energy, one can again
take the limit $n \to 0$ (provided $n$ is even). However, the
calculation of $f$ is found to be easier using the alternative
expression (\ref{eq:alt_free_e}) (see \ref{app:f} for details).
The result of the calculation is
\begin{eqnarray}
\hspace*{-15mm} f &=&  \frac{c}{2\beta} \bra \bra \int\! dh
dh^\prime~ W_{\bxi}(h) W_{\bxi^\prime}(h^\prime) \log\left[1 \!+
\! \tanh(\beta h)\tanh(\beta h^\prime)\tanh(\frac{\beta}{c}
\phi(\bxi \cdot \bxi^\prime)) \right] \ket\ket_{\bxi\bxi^\prime}
\nonumber
\\
\hspace*{-15mm}  && - \frac{c}{2\beta} \bra \bra \log
\cosh[\frac{\beta}{c}\phi(\bxi \cdot \bxi^\prime)]
\ket\ket_{\bxi\bxi^\prime} - \frac{1}{\beta} \bra \int\! dh
~W_{\bxi}(h) \log [2\cosh(\beta h)] \ket_{\bxi} \nonumber \\
\hspace*{-15mm} && - \frac{c}{2\beta} \bra \bra \int\!dh~
W_{\bxi}(h) \log\left[1-\tanh^2(\beta
h)\tanh^2(\frac{\beta}{c}\phi(\bxi \cdot \bxi^\prime))\right]
\ket\ket_{\bxi\bxi^\prime} \label{eq:f_e}
\end{eqnarray}
In deriving (\ref{eq:f_e}), which is reassuringly similar but not
identical to the corresponding expression derived for the RS free
energy of spin glasses with random interactions \cite{MP87}, we
have used $W_{\bxi}(h)$ solving (\ref{eq:W_saddlepoint})
 and being
normalized (for every $\bxi$).

As a simple test of our RS results, one may inspect the limit $c
\to \infty$ for $\phi(x) = x$. Here the interactions are of the
Hopfield type, and, since $p$ is finite, one should recover the
equations describing the Hopfield model away from saturation.
Expansion of the saddle-point equation (\ref{eq:W_saddlepoint}) in
powers of $1/c$, keeping only the $\order((\frac{1}{c})^0)$ terms,
results in
\be
W_{\bxi}(h) = \delta(h - \sum_{\mu} \xi^{\mu} m^{\mu}) \ee Upon
using this expression to calculate the overlaps $m^\mu = \bra
\xi^\mu \int dh W_{\bxi}(h)\tanh(\beta h)\ket_{\bxi} $ one indeed
recovers the saddle-point equations of the Hopfield model away
from saturation. Similarly, (\ref{eq:f_e}) reduces to the correct
corresponding free energy.

\section{Analysis of Phase Transitions}

\subsection{Expansion of the saddle-point equations}

The paramagnetic state $W_{\bxi}(h) = \delta(h)~\forall\bxi$
always solves (\ref{eq:W_saddlepoint}). As the temperature is
lowered from $T=\infty$, we expect other solutions to bifurcate
away from the paramagnetic one. In finite connectivity spin
glasses, phase transitions were found upon expanding the RS free
energy in the order parameters (\ref{eq:orderparams}) near the
paramagnetic phase \cite{VB85}. A similar strategy can be applied
here, following \cite{KS87}.  We assume that, close to the
transition, $\int\!dh~W_{\bxi}(h)h^\ell=\order(\epsilon^\ell)$
with $0<\epsilon\ll 1$, and we expand both sides of
(\ref{eq:saddle_a}) in powers of $\epsilon$. This can be done
self-consistently, for all al orders in $\epsilon$. Identification
of the lowest two orders $\epsilon$ and $\epsilon^2$ is then found
to give, respectively
 \begin{eqnarray} \hspace*{-10mm}
 \int\!dh~W_{\bxi}(h)h =
 c~\bra \int\!dh~W_{\bxi^\prime}(h)h\tanh[\frac{\beta}{c}\phi(\bxi \cdot
\bxi^\prime)]\ket_{\bxi^\prime} \label{eq:eps}
\\
\hspace*{-10mm} \int\!dh~W_{\bxi}(h)h^2 -
\left[\int\!dh~W_{\bxi}(h)h\right]^2= c~\bra
\int\!dh~W_{\bxi^\prime}(h)h^2\tanh^2[\frac{\beta}{c}\phi(\bxi
\cdot \bxi^\prime)]\ket_{\bxi^\prime} \label{eq:epssqr}
\end{eqnarray}
There are hence two types of transitions away from the
paramagnetic (P) state. The first corresponds to the lowest order
being $\epsilon$; such transitions are marked by the appearance of
nonzero solutions of (\ref{eq:eps}). The second type have
$\epsilon^2$ as lowest non-zero order; these transitions are
marked by non-trivial solutions of (\ref{eq:epssqr}) with
$\int\!dh~W_{\bxi}(h)h=0$. We conclude that these transitions are
marked by the highest temperature for which the two $2^p\times
2^p$ matrices $M_{\bxi\bxi^\prime}$ and $Q_{\bxi\bxi^\prime}$,
respectively, have an eigenvalue equal to 1:
\be
M_{\bxi\bxi^\prime} = c p_{\bxi^\prime}
\tanh[\frac{\beta}{c}\phi(\bxi \cdot \bxi^\prime)]  ~~~~~~~~
Q_{\bxi\bxi^\prime} = c p_{\bxi^\prime}
\tanh^2[\frac{\beta}{c}\phi(\bxi \cdot \bxi')] \label{eq:matrices}
\ee It follows from (\ref{eq:RSoverlaps}) that bifurcation of  $M$
eigenvectors corresponds to a transition towards a retrieval state
(R), with nonzero pattern overlaps, whereas
 bifurcation of  $Q$
eigenvectors corresponds to a transition towards a spin-glass
state (SG).

\subsection{Transition Temperatures for Random Patterns}

For independently drawn random patterns, where $p_{\bxi}= 2^{-p}$
for all $\bxi$, both matrices in (\ref{eq:matrices}) are symmetric
and can be diagonalized by exploiting symmetries. The desired
eigenvectors have in fact already been calculated in \cite{LvH}
(in a different context). They are found to be universal for
 all $2^p\times 2^p$ symmetric matrices whose entries depend only on  $\bxi \cdot
 \bxi^\prime$, i.e. $U_{\bxi\bxi^\prime}=U(\bxi\cdot\bxi^\prime)$.
 With each of the $2^p$ index subsets $S\subseteq \{1, 2, \ldots,
 p\}$ one can associate an eigenvector $v_S$, defined as
$v_S(\bxi) = \prod_{\mu \in S} \xi^\mu$. The eigenvector
corresponding to the empty set is defined as $v_{\emptyset}(\bxi)
= 1$ for all $\bxi$. One easily verifies that this gives all
eigenvectors of $U$, with corresponding eigenvalues
\be
\lambda_{S} = \sum_{\bxi} U(\sum_{\nu=1}^p \xi^\nu)\prod_{\mu \in
S} \xi^\mu \label{eq:general_spectrum} \ee The eigenvalues depend
only on the size $|S|$ of the subset. Application of
(\ref{eq:general_spectrum}) to our matrices $M$ and $Q$ gives the
following eigenvalue spectra: \bd \hspace*{-10mm}
 \lambda^M_{S} = c ~\bra
\left[\prod_{\mu \in S}\! \xi^\mu\right]
\tanh[\frac{\beta}{c}\phi(\sum_\nu \xi_\nu)]\ket_{\bxi} ~~~~~~~~
\lambda^Q_{S} = c ~\bra \left[\prod_{\mu \in S}\! \xi^\mu\right]
\tanh^2[\frac{\beta}{c}\phi(\sum_\nu \xi_\nu)]\ket_{\bxi} \ed
Clearly, $\lim_{\beta\to 0}\lambda^M_S=\lim_{\beta\to
0}\lambda^Q_S=0$.
 For monotonically non-decreasing
functions $\phi(x)$, the largest $M$ eigenvalue is found for
$|S|=1$,  whereas the largest $Q$ eigenvalue corresponds to
$S=\emptyset$:
\begin{eqnarray}
\lambda_{\rm max}^M &=& \frac{c}{p} ~\bra (\sum_{\mu}\xi^\mu)
\tanh[\frac{\beta}{c}\phi(\sum_{\nu}\xi^\nu)]\ket_{\bxi}
\label{eq:laMmax}
\\
\lambda_{\rm max}^Q &=& c~\bra
\tanh^2[\frac{\beta}{c}\phi(\sum_{\mu} \xi^\mu)]\ket_{\bxi}
\label{eq:laQmax}
\end{eqnarray}
This leads us to the following equations for the P$\to$R and
P$\to$SG transition lines:
\begin{eqnarray}
{\rm P}\to {\rm R}: &~~~~& \frac{c}{p}~2^{-p}\sum_{n=0}^p
\left(\!\!
\begin{array}{c} p
\\ n
\end{array}\!\!
\right)(p-2n)\tanh[\frac{\beta}{c}\phi(p-2n)]=1 \label{eq:trans1}
\\
{\rm P}\to {\rm SG}: &~~~~&
 c~2^{-p}\sum_{n=0}^p \left(\!\! \begin{array}{c}
p \\ n \end{array} \!\!\right)\tanh^2[\frac{\beta}{c}
\phi(p-2n)]=1 \label{eq:trans2}
\end{eqnarray}
Equations (\ref{eq:trans1},\ref{eq:trans2}) will generally have to
be solved numerically. \vsp

We may finally inspect the limit $c\to \infty$, with $\alpha =p/c$
fixed, where we expect to find a generalization of the results in
\cite{WS91}. The central limit theorem now allows us to replace
$p^{-\frac{1}{2}}\sum_{\mu=1}^p\xi^\mu$ by a zero-average and
unit-variance Gaussian variable, so
(\ref{eq:laMmax},\ref{eq:laQmax}) simplify to (provided the limits
exist, and with $Dy=(2\pi)^{-\frac{1}{2}}e^{-\frac{1}{2}y^2}$):
\begin{eqnarray}
\lim_{p\to\infty}\lambda_{\rm max}^M &=& \frac{1}{\alpha}
\int\!Dy~y \lim_{p\to\infty}\left\{ \sqrt{p}~\tanh[\frac{\alpha
\beta}{p}\phi(y\sqrt{p})]\right\} \label{eq:pinftyR}
\\
\lim_{p\to\infty}\lambda_{\rm max}^Q &=&
\frac{1}{\alpha}\int\!Dy~\lim_{p\to\infty}\left\{p~
\tanh^2[\frac{\alpha\beta}{p}\phi(y\sqrt{p})]\right\}
\label{eq:pinftySG}
\end{eqnarray}

\subsection{Application to Specific Synaptic Kernels}

\begin{figure}[t]
\vspace*{3mm}
\begin{center}
\epsfig{file=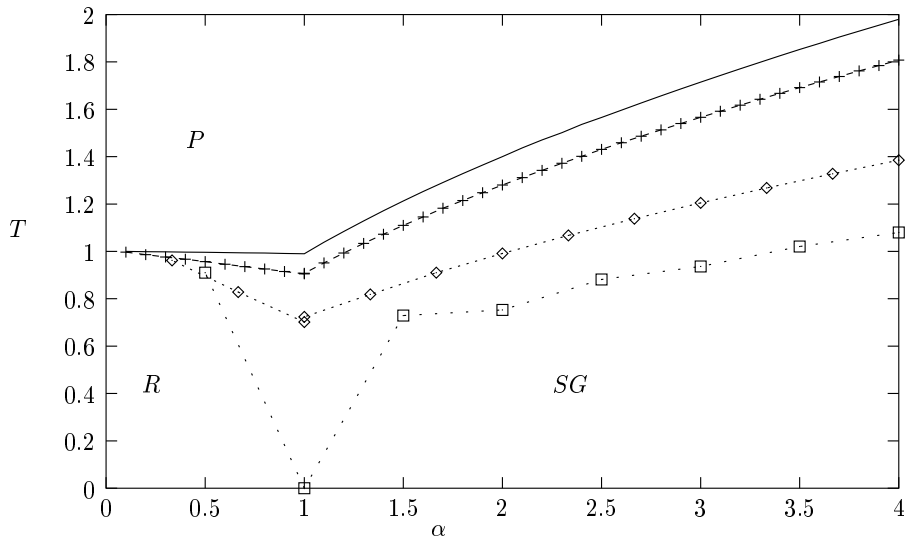, width=12cm} \caption{Phase diagram
of the finite connectivity Hopfield model, where $\phi(x)=x$.
Connected markers give the P$\to$R and P$\to$SG transition
temperatures in the $(\alpha,T)$ plane (where $\alpha=p/c$), for
$c=2$ ($\opensquare$), $c=3$ ($\opendiamond$), $c=10$ ($+$),
whereas the $c=100$ transition is indicated by a solid line. The
transition line for $c=100$ is nearly indistinguishable from the
corresponding $c\to \infty$ line segments $T = 1$ (for $\alpha<1$)
and $T = \sqrt{\alpha}$ (for $\alpha>1$).} \label{fig:fig1}
\end{center}
\end{figure}

\begin{figure}[t]
\vspace*{3mm}
\begin{center}
\epsfig{file=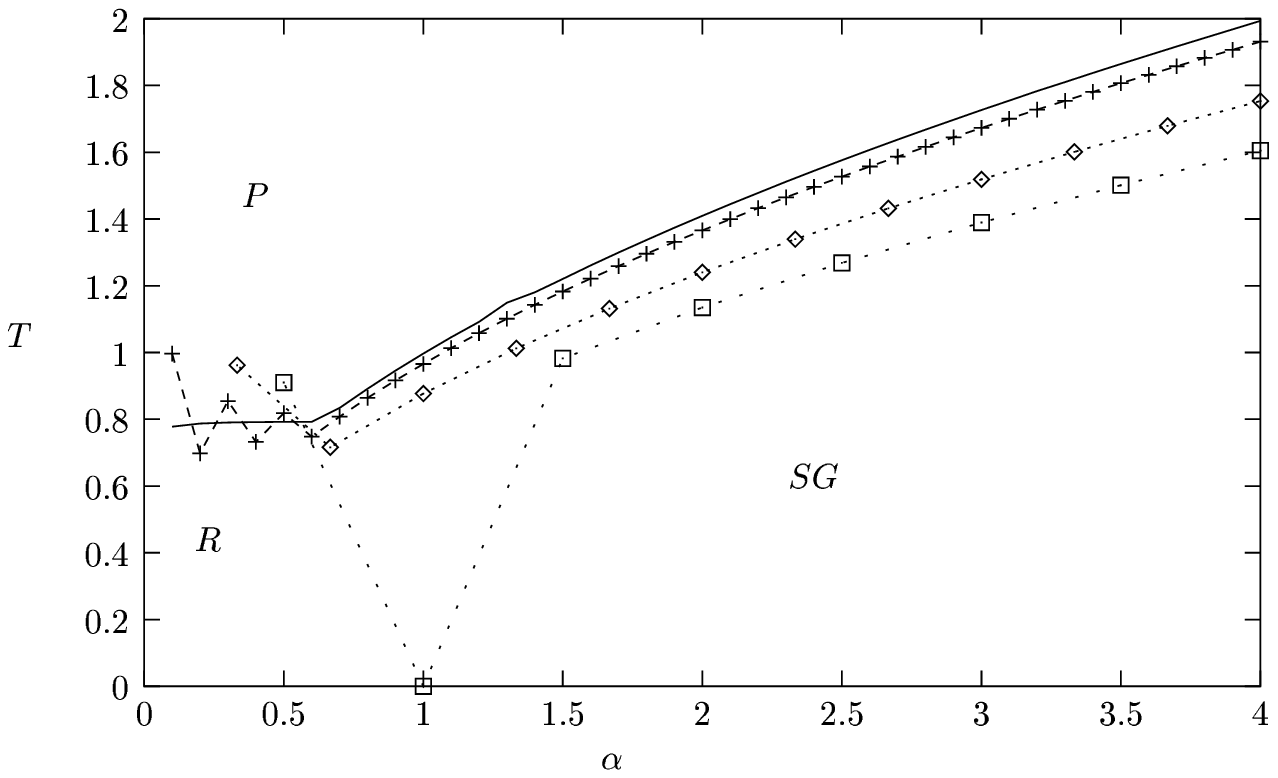, width=12cm} \caption{ Phase
diagram of the finite connectivity clipped Hopfield model, where
$\phi(x)=\sqrt{p}~\sgn(x)$. Connected markers give the P$\to$R and
P$\to$SG transition temperatures in the $(\alpha,T)$ plane (where
$\alpha=p/c$), for $c=2$ ($\opensquare$), $c=3$ ($\opendiamond$),
$c=10$ ($+$), whereas the $c=100$ transition is indicated by a
solid line. The transition line for $c=100$ is again nearly
indistinguishable from the corresponding $c\to \infty$ line
segments, here given by  $T = \sqrt{2/\pi}$ (for
$\alpha<\sqrt{2/\pi}$) and $T = \sqrt{\alpha}$ (for
$\alpha>\sqrt{2/\pi}$). } \label{fig:fig2}
\end{center}
\end{figure}

\begin{figure}[t]
\vspace*{3mm}
\begin{center}
\epsfig{file=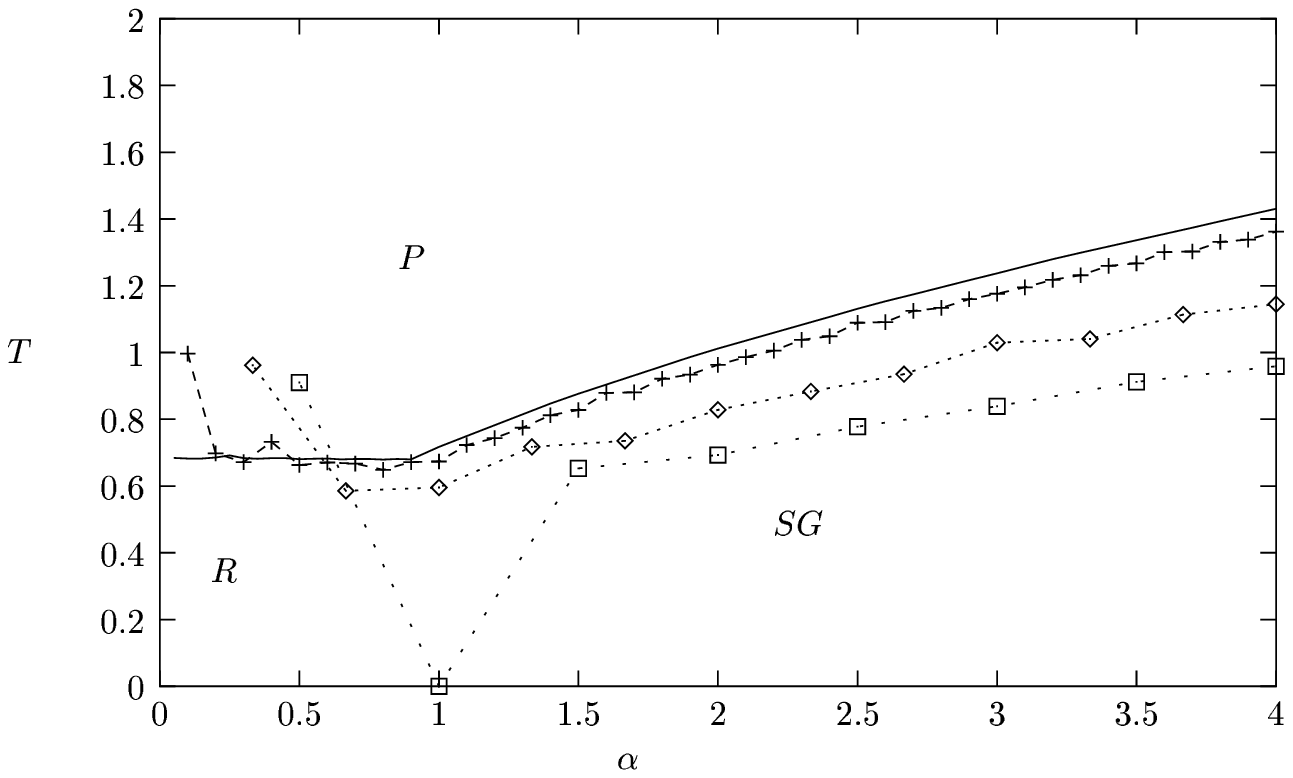, width = 12cm} \caption{
 Phase
diagram of a model intermediate between the previous two, with
$\phi(x)=x$ for $|x|<\sqrt{p}$ and $\phi(x)=\sqrt{p}~\sgn(x)$ for
$|x|>\sqrt{p}$. Connected markers give the P$\to$R and P$\to$SG
transition temperatures in the $(\alpha,T)$ plane, for $c=2$
($\opensquare$), $c=3$ ($\opendiamond$), $c=10$ ($+$), whereas the
$c=100$ transition is indicated by a solid line.
The transition line for $c=100$ is again nearly
indistinguishable from the corresponding $c\to \infty$ line
segments, here given by $T={\rm Erf}(1/\sqrt{2})$ (for small
$\alpha$)  and $T=[\alpha(1-\sqrt{2/\pi e})]^{\frac{1}{2}}$
 (for
large $\alpha$).} \label{fig:fig3}
\end{center}
\end{figure}

We now have to make explicit choices for the kernel $\phi(x)$ in
our expression for the interaction matrix, viz.
$J_{ij}=\frac{c_{ij}}{c}~\phi(\bxi_i\cdot\bxi_j)$. We first
inspect the finite connectivity Hopfield model
$J_{ij}=\frac{c_{ij}}{c}~\bxi_i\cdot\bxi_j$. Here we have
$\phi(x)=x$, and the $c\to\infty$ expressions for the P$\to$R and
P$\to$SG transition temperatures
(\ref{eq:pinftyR},\ref{eq:pinftySG}) reduce to $T_{R}=1$ and
$T_{SG}=\sqrt{\alpha}$ (in accordance with the extreme dilution
results in \cite{WS91}). In addition one finds that for
$c\to\infty$ the effective field distributions $W_{\bxi}(h)$
become Gaussian, with mean $\overline{h}_{\bxi}=\bra
\int\!dh~W_{\bxi^\prime}\tanh(\beta
h)(\bxi\cdot\bxi^\prime)\ket_{\bxi^\prime}$ and variance
$\sigma^2=\alpha \bra\int\!dh~W_{\bxi^\prime}\tanh^2(\beta
h)(\bxi\cdot\bxi^\prime)\ket_{\bxi^\prime}-\overline{h}^2_{\bxi}$,
from which one immediately recovers the RS order parameter
equations of \cite{WS91} (describing states with a single
`condensed' pattern, for the scaling regime $c\to\infty$ with
$c/N\to 0$): \bd m=\int\!Dy~\tanh[\beta(m+y\sqrt{\alpha
q})]~~~~~~~~ q=\int\!Dy~\tanh^2[\beta(m+y\sqrt{\alpha q})] \ed
 The
results of solving (\ref{eq:trans1},\ref{eq:trans2}) numerically
for finite $c$  are shown in figure \ref{fig:fig1}. Note that
$\alpha=p/c\in\{\frac{1}{c},\frac{2}{c},\frac{3}{c},\ldots\}$, so
only the actual markers in figure \ref{fig:fig1} correspond to
physically realizable parameter values (the connecting line
segments just provide a guide to the eye). We observe that for
$c=100$ the transition lines away from the paramagnetic phase are
already nearly identical to those corresponding to $c \to \infty$.
For $c=2$, the system remains in a paramagnetic phase up to $T=0$
for $\alpha=1$. Here, interestingly, for $\alpha = 0.5$ (i.e.
$p=1$) there is a transition to a nonzero retrieval overlap at
finite $T$. This is not what one would find in e.g. a $1-d$ Ising
chain, in which the connectivity is $c=2$, but where there is no
phase transition at finite $T$; the difference is due  to the fact
that in our present model the connectivity equals $2$ only
\emph{on average}.  Note that the value $c=1$ corresponds to the
percolation transition \cite{KS87}. We will turn to the location
of the more complicated  R$\to$SG transition in  a subsequent
section.

Our second application is obtained upon choosing a finite
connectivity version of the so-called clipped Hebbian synapses:
$J_{ij}=\frac{c_{ij}}{c}\sqrt{p}~\sgn(\bxi_i\cdot\bxi_j)$ (the
specific scaling with $\sqrt{p}$ ensures a proper limit
$c\to\infty$). Here $\phi(x)=\sqrt{p}~\sgn(x)$, and the
$c\to\infty$ expressions for the P$\to$R and P$\to$SG transition
temperatures (\ref{eq:pinftyR},\ref{eq:pinftySG}) now reduce to
$T_{R}=\sqrt{2/\pi}$ and $T_{SG}=\sqrt{\alpha}$. The results of
solving (\ref{eq:trans1},\ref{eq:trans2}) numerically for finite
$c$  are shown in figure \ref{fig:fig2}. As before, only the
actual markers in figure \ref{fig:fig2} correspond to physically
realizable parameter values. As was the case for full connectivity
\cite{LvH}, also with finite connectivity one finds, surprisingly
from an information processing point of view, that the differences
between Hebbian and clipped Hebbian synapses are only of a
quantitative nature; limited largely to a rescaling of critical
temperatures.

Our third and final finite connectivity network example is one
which interpolates between Hebbian and clipped Hebbian synapses:
$\phi(x) =x$ (i.e. Hebbian) for $|x|<\sqrt{p}$, and
$\phi(x)=\sqrt{p}~\sgn(x)$ (i.e. clipped Hebbian) for  $|x|\geq
\sqrt{p}$. The value $\sqrt{p}$ is found to be the natural and
most interesting scaling for the cut-off point in this definition.
The $c\to\infty$ expressions for the P$\to$R and P$\to$SG
transition temperatures (\ref{eq:pinftyR},\ref{eq:pinftySG}) now
reduce to $T_{R}={\rm Erf}(1/\sqrt{2})$  and
$T_{SG}=[\alpha(1-\sqrt{2/\pi e})]^{\frac{1}{2}}$.
 The results of
solving (\ref{eq:trans1},\ref{eq:trans2}) numerically for finite
$c$  are shown in figure \ref{fig:fig3}. Again only the
 markers correspond to physically
realizable parameter values.
 The apparent irregularities
in the transition lines are due to the cut-off in the function
$\phi(x)$ at $|\bxi \cdot \bxi^\prime| = \sqrt{p}$.

\section{RS Ground State of the Finite Connectivity Hopfield Model}

For large $\alpha$, one expects replica symmetry to break  at  low
temperatures. In contrast to  full connectivity models, for finite
connectivity models RSB theory is still under development. Here we
will therefore analyze as yet only the $T\to 0$ limit of our RS
equations. Apart from giving exact statements at least for small
$\alpha$, a calculation of the RS ground state entropy leads for
large $\alpha$ to a bound for the location of the RSB transition
in the phase diagram, since the latter must come before the zero
entropy line.

\subsection{RS Order Parameters at $T=0$}

We follow closely the RS ground state analysis carried out for
finite connectivity spin glasses \cite{KS87,MP87}. At $T=0$ there
are no thermal fluctuations, so the effective fields in
(\ref{eq:RSansatz}) are identical to the true local fields. The
latter can take only discrete values, due to the finite number of
connections per spin, hence each $W_{\bxi}(h)$ is a sum of delta
peaks:
\begin{equation}
W_{\bxi}(h) = \sum_{\ell=-\infty}^\infty  K_{\bxi}(\ell)
~\delta(h-\frac{\ell}{c})
 \label{eq:gsdistr}
\end{equation}
Upon taking the limit $\beta \to \infty$ in
(\ref{eq:W_saddlepoint}), using $\lim_{\beta\to\infty}
\beta^{-1}\tanh^{-1}[\tanh(\beta x)\tanh(\beta
y)]=\frac{1}{2}|x+y|-\frac{1}{2}|x-y|$, one verifies by
substitution that (\ref{eq:gsdistr}) is
 indeed a solution.
 Insertion into the right hand side of
 (\ref{eq:W_saddlepoint}) gives (with
 $\overline{\delta}_{nm}=1-\delta_{nm}$):
\begin{eqnarray}
\hspace*{-16mm} W_{\bxi}(h)\! &=& \int\! \frac{dy}{2\pi} e^{-iyh}
\exp\left( c \left\bra \left\{ \sum_{\ell^\prime=1}^{|\bxi \cdot
\bxi^\prime|-1} \Bigl[ K_{\bxi^\prime}(\ell^\prime)
e^{\frac{iy\ell^\prime}{c} \sgn (\bxi \cdot \bxi^\prime)} +
K_{\bxi^\prime}(-\ell^\prime) e^{- \frac{iy\ell^\prime}{c} \sgn
(\bxi \cdot \bxi^\prime)} \Bigr] \right. \right. \right. \nonumber
\\ \hspace*{-16mm} && \hspace*{-13mm} \left. \left. \left. + \sum_{\ell^\prime= |\bxi
\cdot \bxi^\prime|}^{\infty} \Bigl[ K_{\bxi^\prime}(\ell^\prime)
e^{\frac{iy}{c}\bxi \cdot \bxi^\prime} +
K_{\bxi^\prime}(-\ell^\prime) e^{-\frac{iy}{c}\bxi \cdot
\bxi^\prime} \Bigr] - [1-K_{\bxi^\prime}(0)] \right\}
\overline{\delta}_{\bxi \cdot \bxi^\prime\!,0}]
\right\ket_{\bxi^\prime} \right)
\end{eqnarray}
Upon expanding the exponent in the right-hand side, we would
indeed recognize a sum of delta peaks. Integrating of both sides
over an infinitesimally small interval around $h=\ell/c$ leads to
equations for the factors $K^{\pm}_{\bxi}(\ell)$:
\begin{eqnarray}
\hspace*{-25mm} K_{\bxi}(\ell)&=& \lim_{\epsilon \to 0}
\int\!\frac{dz}{\pi z}\sin(z) e^{- \frac{i\ell z}{c\epsilon}}
\exp\left( c \left\bra \left\{ \sum_{\ell^\prime=1}^{|\bxi \cdot
\bxi^\prime|-1} \Bigl[ K_{\bxi^\prime}(\ell^\prime)
e^{\frac{iz\ell^\prime}{c\epsilon} \sgn (\bxi \cdot \bxi^\prime)}
+ K_{\bxi^\prime}(-\ell^\prime) e^{-
\frac{iz\ell^\prime}{c\epsilon} \sgn (\bxi \cdot \bxi^\prime)}
\Bigr] \right. \right. \right. \nonumber\hspace*{-10mm}
\\ \hspace*{-25mm}&& \hspace*{-2mm}\left. \left. \left. + \sum_{\ell^\prime = |\bxi \cdot
\bxi^\prime|}^{\infty} \Bigl[ K_{\bxi^\prime}(\ell^\prime)
e^{\frac{iz}{c\epsilon}\bxi \cdot \bxi^\prime} +
K_{\bxi^\prime}(-\ell^\prime) e^{-\frac{iz}{c\epsilon}\bxi \cdot
\bxi^\prime} \Bigr] - [1-K_{\bxi^\prime}(0)] \right\}
\overline{\delta}_{\bxi \cdot \bxi^\prime,0}
\right\ket_{\bxi^\prime} \right) \label{eq:Keq1}
\end{eqnarray}
For $\epsilon \to 0$, the periodic part of the integral (involving
$z/c\epsilon$) oscillates rapidly and  decouples from the
non-oscillating part. We note that (\ref{eq:Keq1}) is of the form
$\lim_{\epsilon \to 0} \int\!\frac{dz}{\pi
z}\sin(z)f(z/\epsilon)$, where $f(z+2\pi)=f(z)$. The function $f$
may be written as a Fourier series,
$f(z/\epsilon)=\sum_{n=-\infty}^\infty a_n e^{inz/\epsilon}$.
Consequently: \bd \lim_{\epsilon \to 0} \int\!\frac{dz}{\pi
z}\sin(z)f(\frac{z}{\epsilon})= a_0\int\!\frac{dz}{\pi
z}\sin(z)+\sum_{n\neq 0} a_n \lim_{\epsilon \to 0}
 \int\!\frac{dz}{\pi z}\sin(\epsilon z)e^{inz}=a_0
\ed
 Since $a_0= (2\pi)^{-1}\int_{-\pi}^{\pi}\!d\phi~f(\phi)$,
 expression
(\ref{eq:Keq1}) can be simplified to
\begin{eqnarray}
\hspace*{-18mm} K_{\bxi}(\ell) &=& e^{-c \bra
[1-K_{\bxi^\prime}(0)] \overline{\delta}_{\bxi \cdot
\bxi^\prime\!,0} \ket_{\bxi^\prime}} \int_{-\pi}^{\pi}\! \frac{d
\phi}{2\pi} \exp\left( c \left\bra \left\{
\sum_{\ell^\prime=1}^{|\bxi \cdot \bxi^\prime|-1} \Bigl[
K_{\bxi^\prime}(\ell^\prime) + K_{\bxi^\prime}(-\ell^\prime)
\Bigr]\cos(\ell^\prime\phi) \right. \right. \right. \nonumber \\
\hspace*{-18mm}&&\left. \left. \left. \hspace*{20mm} +
\sum_{\ell^\prime = |\bxi \cdot \bxi^\prime|}^{\infty} \Bigl[
K_{\bxi^\prime}(\ell^\prime) +
K_{\bxi^\prime}(-\ell^\prime)]\cos(\bxi \cdot \bxi^\prime
\phi)\right\} \overline{\delta}_{\bxi \cdot \bxi^\prime\!,0}
\right\ket_{\bxi^\prime} \right) \nonumber \\ \hspace*{-18mm}
&&\times \cos\left(- \ell\phi + c \left\bra \left\{
\sum_{\ell^\prime=1}^{|\bxi \cdot \bxi^\prime|-1} \Bigl[
K_{\bxi^\prime}(\ell^\prime) - K_{\bxi^\prime}(-\ell^\prime)
\Bigr]\sin(\ell^\prime\phi)~\sgn(\bxi\cdot\bxi^\prime)\right.\right.\right.\nonumber\\
\hspace*{-18mm}&&\left.\left.\left. \hspace*{20mm}+
\sum_{\ell^\prime = |\bxi \cdot \bxi^\prime|}^{\infty} \Bigl[
K_{\bxi^\prime}(\ell^\prime) -
K_{\bxi^\prime}(-\ell^\prime)]\sin(\bxi \cdot \bxi^\prime
\phi)\right\} \overline{\delta}_{\bxi \cdot \bxi^\prime\!,0}
\right\ket_{\bxi^\prime} \right) \label{eq:gscoeff}
\end{eqnarray}
Solutions to this set of equations can be found numerically, via
iteration.

\subsection{RS Entropy at $T=0$}

We next calculate the RS zero temperature entropy per spin $s_0$,
by expanding the free energy per spin up to order $T$:
\begin{equation}
f = f_{T=0} -s_0 T + \order(T^2)
\end{equation}
Care is needed in taking the temperature derivative of the free
energy, in view of the normalization constraint on the order
parameters $W_{\bxi}(h)$. One generally has
\begin{eqnarray}
 \frac{d f}{d T}
= \left.\frac{\partial f[\ldots]}{\partial T}\right|_{\rm saddle}
+~ \sum_{\bxi^\prime} \int\! dh^\prime \frac{\delta f[\ldots]}
{\delta W_{\bxi^\prime}(h^\prime)}\Biggl|_{\rm saddle}\!\!.
 \frac{\partial W^\star_{\bxi^\prime}(h^\prime)}{\partial T}
 \label{eq:varyf}
\end{eqnarray}
where $W^\star$ denotes the saddle-point of $f[\ldots]$. In
contrast to unconstrained  extremization, the functional
derivative in the right hand side of (\ref{eq:varyf}) need not
vanish, since extremization of $f$ is restricted to the subspace
in which $\int\! dh ~W_{\bxi}(h) = 1$:
\begin{equation}
\frac{\delta f[\ldots]}{\delta W_{\bxi}(h)}  +
\sum_{\bxi^\prime}\lambda_{\bxi^\prime} ~\frac{\delta}{\delta
W_{\bxi}(h)} \left[\int\! dh^\prime~
W_{\bxi^\prime}(h^\prime)-1\right] = 0
\end{equation}
Here $\{\lambda_{\bxi^\prime}\}$ are Lagrange multipliers. As a
consequence, we find $\delta f[\ldots]/\delta W_{\bxi}(h) =
\lambda_{\bxi}$ at the saddle point (i.e. independent of the field
$h$), and hence
\begin{eqnarray}
 \frac{d f}{d T}
&=& \left.\frac{\partial f[\ldots]}{\partial T}\right|_{\rm
saddle} + \sum_{\bxi^\prime} \lambda_{\bxi^\prime}
\frac{\partial}{\partial T} \int\! dh^\prime~
W^\star_{\bxi^\prime}(h^\prime)=\left.\frac{\partial
f[\ldots]}{\partial T}\right|_{\rm saddle}
\end{eqnarray}
For the purpose of calculating $s_0$,  we may thus simply insert
the $T=0$ saddle-point (\ref{eq:gsdistr},\ref{eq:gscoeff}) into
our expression (\ref{eq:f_e}) for the free energy, giving:
\begin{eqnarray}
\hspace*{-20mm} s_0 &=& \lim_{\beta \to \infty}
\beta^2\frac{\partial}{\partial\beta} \left\bra \!\left\bra~
\frac{c}{2\beta} \sum_{\ell,\ell^\prime\neq 0} K_{\bxi}(\ell)
K_{\bxi^\prime}(\ell^\prime) \log[1+ \tanh(\frac{\beta \ell}{c})
\tanh(\frac{\beta \ell^\prime}{c}) \tanh(\frac{\beta}{c} \bxi
\cdot \bxi^\prime)] \right.\right. \nonumber
\\ \hspace*{-20mm} && \hspace*{-8mm}\left.\left.
-\frac{c}{2\beta} \sum_{\ell\neq 0} K_{\bxi}(\ell)\left[ \log[1+
\tanh(\frac{\beta \ell}{c}) \tanh(\frac{\beta}{c} \bxi \cdot
\bxi^\prime)] + \log[1- \tanh(\frac{\beta \ell}{c})
\tanh(\frac{\beta}{c} \bxi \cdot \bxi^\prime)]\right]
\right.\right. \nonumber \\ \hspace*{-20mm}&&
\hspace*{15mm}\left.\left. -\frac{c}{2\beta} \log
\cosh(\frac{\beta}{c}\bxi\cdot \bxi^\prime)
 - \frac{1}{\beta}  \sum_{\ell\neq 0}
K_{\bxi}(\ell) \log[2\cosh(\frac{\beta \ell}{c})]~
\right\ket\!\right\ket_{\bxi\bxi^\prime} \nonumber
\\
\hspace*{-20mm} & =& \frac{c}{2}\log 2~\left\bra \left\bra
\sum_{\ell=0}^{|\bxi \cdot \bxi^\prime|-1}\left\{ K_{\bxi}(-|\bxi
\cdot \bxi^\prime|) + K_{\bxi}(|\bxi  \cdot \bxi^\prime|)
-[K_{\bxi}(\ell) + K_{\bxi}(-\ell)][1-\frac{1}{2}\delta_{\ell,0}]
\right\} \right. \right. \nonumber \\ \hspace*{-20mm} &&
\hspace*{30mm}\left. \left. \times ~[K_{\bxi^\prime}(\ell) +
K_{\bxi^\prime}(-\ell)] [1-\frac{1}{2}\delta_{\ell,0}]
\overline{\delta}_{\bxi \cdot \bxi^\prime,0} \right\ket
\right\ket_{\bxi\bxi^\prime} + \log 2\bra K_{\bxi}(0) \ket_{\bxi}
\nonumber \\ \hspace*{-20mm} && + \frac{c}{4} [2\log2 - \log3]
~\left\bra \left\bra ~[K_{\bxi}(-|\bxi \cdot \bxi'|) +
K_{\bxi}(|\bxi \cdot \bxi'|)] \right. \right. \nonumber \\
\hspace*{-20mm} && \hspace*{50mm}\times \left. \left.
[K_{\bxi'}(-|\bxi \cdot \bxi'|) + K_{\bxi'}(|\bxi \cdot
\bxi'|)]~\overline{\delta}_{\bxi \cdot \bxi^\prime,0} \right\ket
\right\ket_{\bxi\bxi^\prime} \label{eq:gsentrop}
\end{eqnarray}
This result is reminiscent of the corresponding expression  found
for spin glasses in \cite{Al02}.

\subsection{Zero Temperature Phase Transitions}

\begin{figure}[t]
\vspace*{3mm}
\begin{center}
 \epsfig{file=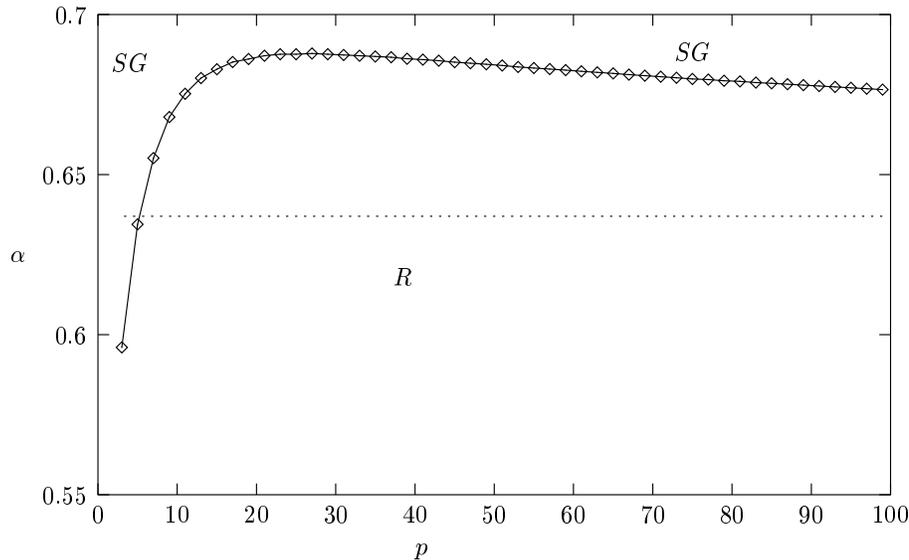, width= 12cm}
 \caption{Zero temperature phase diagram of
the finite connectivity Hopfield model, in RS approximation and
for odd $p$ (note: $p=\alpha c$). Connected diamonds
$(\diamondsuit)$: the R$\to$SG transition line. Dashed horizontal
line: the $c\to\infty$ location for the R$\to$SG line,
$\alpha_c=2/\pi$, as known from  \cite{WS91}.}
\label{fig:zerotemp}
\end{center}
\end{figure}

We next try to solve the coupled equations (\ref{eq:gscoeff}), for
randomly drawn patterns, at zero temperature. We know that in the
case of perfect retrieval of a pattern $\mu$, half of the
sublattice magnetizations (the ones with $\xi^\mu = 1$) must be
$1$ and the other half must be $-1$. At $T=0$ all spins align with
their local fields, so it follows from (\ref{eq:gsdistr}) that
$K_{\bxi}(\ell)=K(\xi^\mu\ell)$ for some set of non-negative
numbers $K(\ell)$. We will now choose the property
$K_{\bxi}(\ell)=K(\xi^\mu\ell)$ for all $\bxi\in\{-1,1\}$ and all
integer $\ell$ as an {\em ansatz}. Upon restricting ourselves for
simplicity to odd values of $p$ , our equations (\ref{eq:gscoeff})
subsequently simplify to
\begin{eqnarray}
K(\ell) &=& e^{-c[1-K(0)]} \int_{-\pi}^{\pi}\! \frac{d\phi}{2\pi} ~e^{cA(\phi,
\{K\})}\cos[-\ell\phi + cB(\phi, \{K\})]
\label{eq:finalKeq}
\end{eqnarray}
with the abbreviations
\begin{eqnarray}
\hspace*{-15mm}
 A(\phi, \{K\})
&=& 2^{-(p-1)}\!\sum_{m=(p+1)/2}^p \left(\! \begin{array}{c} p \\
m
\end{array}\! \right)  \sum_{\ell=1}^{2m-p-1}[K(\ell) + K(-\ell)]\cos(\ell\phi)
\nonumber \\ \hspace*{-15mm} && + 2^{-p} \sum_{m=0}^p \left(
\!\begin{array}{c} p \\ m
\end{array}\! \right) \cos[(p-2m)\phi]
\!\sum_{\ell=|p-2m|}^{\infty}[K(\ell) + K(-\ell)]
\\
\hspace*{-15mm} B(\phi, \{K\}) &=& 2^{-p}\sum_{m=0}^p \frac{p-
2m}{p} \left(\!
\begin{array}{c} p \\ m
\end{array}\! \right) \sin[(p-2m)\phi]
\!\sum_{\ell=|p-2m|}^{\infty} [K(\ell) - K(-\ell)]
\end{eqnarray}
The result of solving (\ref{eq:finalKeq}) numerically, and
subsequently detecting the R$\to$SG  transition (which is found to
be first order) is shown in figure \ref{fig:zerotemp} in the
$(p,\alpha)$ plane.
 The R$\to$SG transition line should for $c\to\infty$ (i.e. $p\to\infty$ with fixed $\alpha$)
 approach the value $\alpha_c=2/\pi$ ( $\approx 0.637$), in accordance with
 \cite{WS91}; the latter limit is shown in the figure as a horizontal dashed
 line, which is indeed seen to be approached by our R$\to$SG line as $p\to\infty$.
 As usual for the present type of system, we
 concentrated on bifurcation points.
 Due to the extensive energy barriers between the ergodic sectors in attractor neural
 networks, on the important time-scales local stability is more important than thermodynamic stability, so the
 thermodynamic transition lines (based on comparing values of free energies)
 are not relevant.
Numerical evaluation of the free energy of the candidate solution
in fact reveals that the bifurcation lines coincide with the
thermodynamic transitions.

 Figure \ref{fig:zerotemp} shows re-entrance phenomena. It
should be noted that an increase of $p$ for fixed $\alpha$ implies
a simultaneous  increase of the number of stored patterns  (with a
detrimental effect on recall quality) and of the connectivity $c$
(expected to have a positive effect on recall quality); the
non-monotonic appearance of the R$\to$SG line reflects the
competition of these opposite tendencies.

Finally,  for the present type of zero temperature solution, our
expression for the RS zero temperature entropy (\ref{eq:gsentrop})
 reduces to
\begin{eqnarray}
\hspace*{-20mm}\frac{ s_0}{\log 2} &=&  K(0)+c ~ 2^{-p}
\sum_{m=(p+1)/2}^p \left(\!\!
\begin{array}{c} p \\ m \end{array} \!\!\right) \left\{
\sum_{\ell=0}^{2m-p-1} \Biggl[K(2m-p)+K(p-2m) \right. \nonumber \\
\hspace*{-20mm}&& \hspace*{33mm} \left.  - [K(\ell) +
K(-\ell)][1-\frac{1}{2}\delta_{\ell,0}] \Biggr] [K(\ell) +
K(-\ell)][1-\frac{1}{2}\delta_{\ell,0}] \right\} \nonumber \\
\hspace*{-20mm}&&+ c~2^{-p}  (1 - \frac{\log3}{2\log 2} )
\sum_{m=(p+1)/2}^p \left(\!\! \begin{array}{c} p \\ m
\end{array}\!\! \right)[K(2m-p) + K(p-2m)]^2
\end{eqnarray}
Numerical evaluation of this expression reveals that the RS
entropy at $T=0$ is never zero, for any combination of $\alpha$
and $p$. This at first sight somewhat surprising result is in fact
in accordance with a similar observation made recently for finite
connectivity spin-glasses in \cite{Al02}. It does, however, not
imply that replica symmetry is not broken at low temperatures
(which we know must happen for $c\to\infty$ \cite{WS91}).

\section{Discussion}

In this paper we have shown how the mathematical framework of
finite connectivity spin glass replica theory (involving order
parameter functions in replica space, rather than matrices) can be
combined with the concept of sub-lattices in order to solve a
large class of finite connectivity Hopfield-type attractor neural
network models near saturation. So far such networks appear to
have been studied only in the two technically simpler regimes of
full connectivity and so-called extreme dilution, both of which
involve a diverging number of bonds per spin $c$ in the
thermodynamic limit. We have restricted ourselves to a replica
symmetric (RS) analysis. The replica symmetric theory is found to
lead to $2^p$ coupled integral equations (\ref{eq:W_saddlepoint}),
whose solutions give the effective local field distributions in
each of the $2^p$ sublattices of the system. Here $p$ denotes the
number of stored patterns. In the limit $c\to\infty$ our equations
are shown to reduce to the theory of so-called extremely diluted
attractor neural networks, as expected. For $T\to 0$ and below a
certain critical value for $\alpha$ one should expect  replica
symmetry no longer to hold (it has been shown for $c\to\infty$ in
\cite{WS91}); going beyond replica symmetry would require
extending the theory to include one or more steps of replica
symmetry breaking, similar to the finite connectivity spin glass
calculations  in \cite{WS88,DG89,LG90,GD90,Mo98,PT02}.

As is usual for attractor neural networks near saturation, our
phase diagrams exhibit three phases: a paramagnetic phase (P), a
retrieval phase (R), and a spin-glass phase (SG). We have
calculated, for arbitrary values of the number of bonds $c$ per
spin and a large class of Hebbian type synaptic kernels, the
(second order)  P$\to$R and P$\to$SG transition lines in the
$(\alpha,T)$ phase diagram. For the main member of our model
class, the finite connectivity Hopfield model, we have also
calculated the (second order) R$\to$SG transition line at $T=0$.
We find that the values of the average connectivity $c$ needed for
the system to function as an attractor neural network are
surprisingly small, barely exceeding the percolation threshold,
even for clipped Hebbian synapses (where each bond carries only
one bit of information). Figures \ref{fig:fig1} to \ref{fig:fig3},
for instance, underline that already for $c=3$ the phase diagrams
differ only in a relatively modest sense from those corresponding
to $c\to\infty$. This underlines the robustness of recurrent
neural networks as information processing systems.

Apart from the obvious extension of our present work, the
inclusion of RSB solutions of our saddle-point equations and
calculation of AT lines \cite{AT}, it would also be an interesting
challenge to attempt a dynamical solution. Within the generating
functional analysis formalism of \cite{DeDom78} this would for
finite connectivity systems involve an effective single spin
problem, with order parameters describing single-spin path
probabilities and the effect on these path probabilities of
single-spin external field perturbations.

\section*{Acknowledgement}

One of the authors (BW) gratefully acknowledges helpful
discussions with Prof JRL De Almeida.

\clearpage
\appendix

\section{Calculation of the RS Free Energy}
\label{app:f}

In this section we calculate the RS free energy per spin, upon
inserting (\ref{eq:RSansatz}) into  (\ref{eq:alt_free_e}). In
doing so we will use the short-hands: \bd
\gamma_m=\int_{-\pi/2}^{\pi/2}\!\frac{d\phi}{\pi}~\log[\cos(\phi)].\cos(2m\phi)~~~~~~~~~m=0,\pm
1,\pm 2,\ldots \ed They obey: \be \sum_{m=-\infty}^\infty
\gamma_m~ e^{2am}=\log\cosh(a) \label{eq:gamma_relation} \ee
Taking the limit $n\to 0$ is again found to impose the need to
restrict ourselves to even values of $n$. One then obtains, after
some algebra, for the entropic term in (\ref{eq:alt_free_e}):
\begin{eqnarray}
\hspace*{-20mm} \lim_{n\to 0}\frac{1}{\beta n} \bra \sum_{\bsigma}
P_{\bxi}(\bsigma) \log P_{\bxi}(\bsigma) \ket_{\bxi}
 &=&\lim_{n\to 0} \frac{1}{\beta
n} \sum_{m=-\infty}^{\infty} \int_0^{2\pi}\! \frac{d\hat{m}}{2\pi}
e^{i\hat{m}m} \int\! dh~e^{\beta
hm}\left[\frac{\cos(\hat{m})}{\cosh(\beta h)}\right]^n\nonumber
\\
&&\hspace*{9mm} \times~\bra W_{\bxi}(h) \log \left[ \int\!
dh^\prime~ W_{\bxi}(h^\prime)\frac{e^{\beta h^\prime
m}}{[2\cosh(\beta h^\prime)]^n}\right]\ket_{\bxi} \nonumber\\
 &&\hspace*{-30mm}=
\frac{1}{\beta}\sum_{m=-\infty}^\infty\gamma_m \int\!dh~\bra
W_{\bxi}(h)e^{2\beta
hm}\log\left[\int\!dh^\prime~W_{\bxi}(h^\prime)e^{2\beta h^\prime
m}\right]\ket_{\bxi} \nonumber\\ &&
 -\frac{1}{\beta}\int\!dh~\bra W_{\bxi}(h)\log [2\cosh(\beta
h)]\ket_{\bxi} \label{eq:entropic_part}
\end{eqnarray}
Similarly, for the energetic term $U$ in (\ref{eq:alt_free_e}) one
finds:
\begin{eqnarray}
\hspace*{-22mm} -\lim_{n\to 0}\frac{c}{2\beta n} \bra \bra
\sum_{\bsigma\bsigma^\prime} P_{\bxi}(\bsigma)
P_{\bxi^\prime}(\bsigma') \left[ e^{\frac{\beta}{c}\phi(\bxi \cdot
\bxi^\prime)(\bsigma \cdot \bsigma^\prime)}-1\right]
\ket\ket_{\bxi\bxi^\prime}\nonumber\\  & &\hspace*{-65mm}
=-\frac{c}{2\beta}\int\!dh dh^\prime~\bra\bra
W_{\bxi}(h)W_{\bxi^\prime}(h^\prime)\log\left[
\sum_{\sigma\sigma^\prime}e^{\frac{\beta}{c}\phi(\bxi\cdot\bxi^\prime)\sigma
\sigma^\prime +\beta[h\sigma+h^\prime\sigma^\prime]}\right]
\ket\ket_{\bxi\bxi^\prime} \nonumber\\ && \hspace*{-60mm}
+\frac{c}{\beta}\int\!dh~\bra W_{\bxi}(h)\log \cosh(\beta
h)\ket_{\bxi} \nonumber
\\
&&\hspace*{-97mm} =-\frac{c}{2\beta}\int\!dh dh^\prime~\bra\bra
W_{\bxi}(h)W_{\bxi^\prime}(h^\prime)\log\left[1+\tanh(\beta
h)\tanh(\beta
h^\prime)\tanh(\frac{\beta}{c}\phi(\bxi\cdot\bxi^\prime))\right]\ket\ket_{\bxi\bxi^\prime}
\nonumber\\ && \hspace*{-60mm}
-\frac{c}{2\beta}\bra\bra\log\cosh[\frac{\beta}{c}\phi(\bxi\cdot\bxi^\prime)]\ket\ket_{\bxi\bxi^\prime}
\label{eq:energetic_term}
\end{eqnarray}
We use (\ref{eq:saddle_a}) to simplify the entropic term
(\ref{eq:entropic_part}), and add the latter to the energetic term
(\ref{eq:energetic_term}) to obtain the following expression for
the  RS free energy per spin:
\begin{eqnarray*}
\hspace*{-20mm}
 f_{\rm RS}&=&-\frac{1}{\beta}\int\!dh~\bra
W_{\bxi}(h)\log[2\cosh(\beta h)]\ket_{\bxi}
-\frac{c}{2\beta}\bra\bra\log\cosh[\frac{\beta}{c}\phi(\bxi\cdot\bxi^\prime)]\ket\ket_{\bxi\bxi^\prime}\nonumber
\\
\hspace*{-20mm} && \hspace*{-5mm}
-\frac{c}{2\beta
}\int\!dhdh^\prime~\bra\bra W_{\bxi}(h)W_{\bxi^\prime}(h^\prime)
\log\left[1+\tanh(\beta h)\tanh(\beta
h^\prime)\tanh(\frac{\beta}{c}\phi(\bxi\cdot\bxi^\prime))\right]\ket\ket_{\bxi\bxi^\prime}
\nonumber
\\
\hspace*{-20mm} &&
\hspace*{-10mm}
+\frac{c}{\beta}\sum_{m=-\infty}^\infty \gamma_m
\int\!dhdh^\prime~\bra\bra
W_{\bxi}(h)W_{\bxi^\prime}(h^\prime)e^{2\beta hm}\left[
e^{2m\tanh^{-1}[\tanh(\beta
h^\prime)\tanh(\frac{\beta}{c}\phi(\bxi\cdot\bxi^\prime))]}-1\right]\ket\ket_{\bxi\bxi^\prime}
\end{eqnarray*}
At this stage we may use relation (\ref{eq:gamma_relation}) to
carry out the summation over $m$. After some further
re-arrangement of terms, and application of the simple identity
$\cosh[\tanh^{-1}(y)]=(1-y^2)^{-1/2}$, one then arrives at the
final result (\ref{eq:f_e}).

\end{document}